# Order parameter dynamics of a Bose-Einstein condensate of exciton-polaritons in semiconductor microcavities

I. Shelykh[1,2], F.P. Laussy[1], A.V. Kavokin[1] and G. Malpuech[1]

[1]LASMEA, UMR 6602 CNRS, Université Blaise-Pascal, 24, av des Landais, 63177, Aubière, France.

[2]St. Petersburg State Polytechnical University, 29, Politechnicheskaya, 195251, St-Petersburg, Russia.

It is shown theoretically that Bose condensation of spin-degenerated exciton-polaritons results in spontaneous buildup of the linear polarization in emission spectra of semiconductor microcavities. The linear polarization degree is a good order parameter for the polariton Bose condensation. If spin-degeneracy is lifted, an elliptically polarized light is emitted by the polariton condensate. The main axis of the ellipse rotates in time due to self-induced Larmor precession of the polariton condensate *pseudospin*. The polarization decay time is governed by the dephasing induced by the polariton-polariton interaction and strongly depends on the statistics of the condensed state.

Bose-condensation of exciton-polaritons (polaritons) in microcavities [1] is now in the focus of experimental and theoretical research. Possessing an extremely light effective mass (of the order of $10^{-4}\,m_0$), polaritons may condense even at room temperature provided that their lifetime is sufficiently long with respect to the characteristic thermalization time [2]. In realistic microcavities, polaritons may have a strongly non-equilibrium distribution in reciprocal space and their condensation is dramatically dependent on their relaxation dynamics. A clear experimental criterion for the condensation has been a subject for debate during recent years [3-4]. Stimulated scattering of polaritons to their ground state, narrowing of the photoluminescence line and nonlinear dependence of the emission intensity have been reported [5] but can only certify of an accumulation of a large number of polaritons in their ground state. The information about the statistics of the polaritons in the ground state can be obtained from the second order coherence $g_2(0)$ measured with a Hanbury Brown-Twiss setup [3]. However, strictly speaking, Bose condensation is a phase transition linked with the spontaneous symmetry breaking of gauge invariance, that is, with appearance of a well defined phase in the system, which cannot be evidenced by the above mentioned experiments. Such a phase transition manifests itself in the spontaneous appearance of a non-zero, long living *order parameter* of the condensate which can be interpreted as an average complex amplitude of the polariton field inside the cavity [6,7]. The observation of such an order parameter is difficult if the measurements are performed on a purely circularly polarized polariton state. On the other hand, if two spin-polarized condensates coexist, their interferences give rise to a very particular temporal dependence of linear polarization of the emitted light. The quantum properties of the light emitted by a polariton condensate have been addressed theoretically in a number of publications [6, 7-12]. All these works, however, ignored the polarization of cavity modes. Recent experiments have shown that the energy relaxation of polaritons is polarization-dependent [13] and that spin-dynamics in microcavities is extremely rich and complicated [14, 15].

In this Letter we propose a simple experimental method to evidence the appearance and survival of the order parameter of a condensate made of interacting polaritons. We show that a spontaneous symmetry breaking in an ensemble of polaritons manifests itself in a dramatic change of the linear polarization degree of the light emitted by the cavity and that the lifetime of this polarization depends strongly on the nature of the polariton state. If the ground state of the system is spin-degenerated, appearance of the order parameter in the condensate leads to the spontaneous buildup of a linear polarization whose in-plane orientation is constant in time, but randomly changes from experiment to experiment in isotropic system. But the spin-degeneracy can also be significantly lifted by fluctuations feeding spin-up and spin-down condensates with unequal populations. For classical particles these fluctuations would yield a mean unbalance of $\sqrt{n_{0,\uparrow}+n_{0,\downarrow}}$ particles between the two condensates with $n_{0,\uparrow}$ spin-up and $n_{0,\downarrow}$ spin-down particles and consequently the circular polarization degree $\rho_c \equiv (n_{0,\uparrow}-n_{0,\downarrow})/(n_{0,\uparrow}+n_{0,\downarrow})$ would vanish like the inverse square root of the occupation number. However, because of stimulation, the probability to reach one condensate or the other depends on respective populations in such a way as to strengthen the more populated state, leading to possibly highly degenerated configurations. The probability for a particle to join the condensate with $n_{0,\uparrow\downarrow}$ particles is $p_{\uparrow\downarrow} = (n_{0,\uparrow\downarrow}+1)/(n_{0,\uparrow}+n_{0,\downarrow}+2)$. This yields $\langle|\rho_c|\rangle = (2+n_0)/(2+2n_0)$ which is approximately ½ for large values of $n_0 \equiv n_{0,\uparrow}+n_{0,\downarrow}$, which corresponds to an elliptically polarized light.

We consider an isotropic microcavity pumped out of resonance and incoherently by an unpolarized *cw* light-source. We do not discuss the dynamics of the polariton condensate formation which has already been described elsewhere [6]. Our goal is to describe the time evolution and dephasing of the condensate (and therefore of the linear polarization) versus its coherence degree in the stationary regime. In this regime, a balance between incoming and out-going particles is reached in the ground state, and the energy distribution function of polaritons does not change in time.

The Hamiltonian we consider has the general form of the interaction Hamiltonian for spin-polarized exciton-polaritons. It retains all interactions essential for the effect we discuss and neglects coupling to the excited states, the lifetime, scattering towards spin-forbidden ("dark") exciton states, radiative decay and spin-lattice relaxation:

$$H = \varepsilon\left(a_\uparrow^+ a_\uparrow + a_\downarrow^+ a_\downarrow\right) + W_1\left(a_\uparrow^+ a_\uparrow^+ a_\uparrow a_\uparrow + a_\downarrow^+ a_\downarrow^+ a_\downarrow a_\downarrow\right) + W_2 a_\uparrow^+ a_\uparrow a_\downarrow^+ a_\downarrow \; , \tag{1}$$

where $a_{\uparrow,\downarrow}, a_{\uparrow,\downarrow}^+$ are annihilation and creation operators for ground state polaritons with spins up or down, $\varepsilon$ is the energy of the polariton ground state, $W_1$ and $W_2$ are interaction constants for polaritons having parallel or antiparallel spins, respectively. The Hamiltonian (1) describes dephasing of the polaritons in the condensate caused by their interactions. It neglects however the dephasing induced by spontaneous scattering of exciton-polaritons from the excited states to the condensate (further referred to as spontaneous dephasing). Its rate is given by $D \approx \dfrac{\Gamma_0}{2n_0}$ [12] where $\Gamma_0$ is the radiative broadening. When $n_0$ is large, this dephasing becomes negligible compared with the energy shifts and the energy broadenings induced by the polariton-polariton interaction, as shown below.

The intensities $I_+$ and $I_-$ of the circular polarized components of light emitted by the polariton condensate under consideration are proportional to $\langle a_\uparrow^+ a_\uparrow \rangle = n_\uparrow$ and $\langle a_\downarrow^+ a_\downarrow \rangle = n_\downarrow$ respectively, while the linear-polarized components of the emitted light $I_\leftrightarrow, I_\updownarrow$ are

$$I_\leftrightarrow \sim \frac{1}{2}\left[\langle a_\uparrow^+ a_\uparrow \rangle + \langle a_\downarrow^+ a_\downarrow \rangle\right] + \mathrm{Re}\langle a_\uparrow a_\downarrow^+ \rangle = \frac{1}{2}\left[n_\uparrow + n_\downarrow\right] + S_x , \qquad (2)$$

$$I_\updownarrow \sim \frac{1}{2}\left[\langle a_\uparrow^+ a_\uparrow \rangle + \langle a_\downarrow^+ a_\downarrow \rangle\right] - \mathrm{Re}\langle a_\uparrow a_\downarrow^+ \rangle = \frac{1}{2}\left[n_\uparrow + n_\downarrow\right] - S_x , \qquad (3)$$

The operator of linear polarization $\hat{S} = a_\uparrow a_\downarrow^+$ determines the in-plane components $S_{x,y}$ of the polariton pseudospin $S_x + iS_y = S = \langle \hat{S} \rangle$ and governs the linear polarization degree of the emitted light:

$$\rho_L = 2|S|/(n_\uparrow + n_\downarrow). \qquad (4)$$

In Heisenberg representation, the dynamics of $\hat{S}$ is:

$$\frac{d\hat{S}}{dt} = \frac{i}{\hbar}\left[\hat{S}; H\right] = \frac{i}{\hbar}\left[V\left(\hat{n}_\downarrow - \hat{n}_\uparrow - 1\right)\right]\hat{S} \qquad (5)$$

where $V = 2W_1 - W_2$. For exciton-polaritons in a microcavity, $V \approx 6xE_b a_b^2 / L^2$ [8], where $x$ is the exciton fraction of the polariton, $E_b$ is the exciton binding energy, $a_b$ is the two-dimensional exciton Bohr radius and $L^2$ is the area of the cavity. In CdTe microcavities with a lateral size 10 μm, $V$ is about 10 neV. The pseudospin temporal dependence reads:

$$S(t) = e^{-iVt/\hbar} Tr\left[\exp\left[\frac{iVt}{\hbar}\left(\hat{n}_\downarrow - \hat{n}_\uparrow\right)\right]\hat{S}(0)\hat{\rho}\right] = e^{-iVt/\hbar} Tr\left[\exp\left[-\frac{iVt}{\hbar}\hat{n}_\uparrow\right]a_\uparrow \rho_\uparrow\right] \cdot Tr\left[\exp\left[\frac{iVt}{\hbar}\hat{n}_\downarrow\right]a_\downarrow^+ \rho_\downarrow\right],$$

(6)

where $\rho$, the density matrix of the system, is time-independent in the Heisenberg representation. $S(t)$ describes variation of the intensity of light emitted by the cavity in a given linear polarization. We have assumed that the two condensates are not correlated which allows to factorise the density matrix as $\rho = \rho_\uparrow \otimes \rho_\downarrow$. The initial in-plane pseudospin reads:

$$S(0) = \alpha_\uparrow \alpha_\downarrow^* \quad (7)$$

where $\alpha_{\uparrow\downarrow} = Tr\left[a_{\uparrow\downarrow}\rho_{\uparrow\downarrow}\right]$ is the order parameter of each circularly polarized condensate. According to (4) and (7), the appearance of the linear polarization in the condensate can be observed only if an order parameter builds up for each of the circularly polarized components. The measurement of the circularly polarized emission gives access to $n_{\uparrow\downarrow}$ which combined with the measurement of the linear polarization degree gives a measurement of the order parameter. Note that the superposition of two states with a Poisson distribution but no well defined phase (so-called randomly phased coherent states) does not lead to an in-plane polarization.

In what follows we compute the time dependence of the in-plane pseudospin versus the coherence degree of the individual condensates. We use the Glauber-Sudarshan representation of the density matrix, $\rho_{\uparrow,\downarrow} = \int |\alpha_{\uparrow,\downarrow}\rangle P(\alpha_{\uparrow,\downarrow})\langle \alpha_{\uparrow,\downarrow}| d\alpha_{\uparrow,\downarrow}$, with $\alpha$ characterizing the coherent state $|\alpha\rangle$ (with a given amplitude and phase) [17]. From this definition one obtains:

$$S(t) = e^{-iVt/\hbar} \int P_\uparrow(\alpha_\uparrow)\langle\alpha_\uparrow|e^{-\frac{iVt}{\hbar}\hat{n}_\uparrow}a_\uparrow|\alpha_\uparrow\rangle d\alpha_\uparrow \cdot \int P_\downarrow(\alpha_\downarrow)\langle\alpha_\downarrow|e^{\frac{iVt}{\hbar}\hat{n}_\downarrow}a_\downarrow^+|\alpha_\downarrow\rangle d\alpha_\downarrow \quad (8)$$

The initial coherence degree in the individual condensates is given by:

$$\chi_{\uparrow\downarrow} = \frac{|\alpha_{\uparrow\downarrow}|^2}{n_{\uparrow\downarrow}} = \frac{n_{\uparrow\downarrow,coh}}{n_{\uparrow\downarrow,coh} + n_{\uparrow\downarrow,th}} = \sqrt{2 - g_{2,\uparrow\downarrow}(0)}, \quad (9)$$

where $g_{2,\uparrow\downarrow}(0)$ is the second order coherence of the individual condensates, $n_{\uparrow,th}, n_{\downarrow,th}$ are the average numbers of spin-up and spin-down polaritons in the thermal fraction, $n_{\uparrow\downarrow,coh} = |\alpha_{\uparrow\downarrow}|^2$ are the average numbers of spin-up and spin-down polaritons in the coherent fraction. The coherence degree varies between 0 (thermal state) and 1 (coherent state). Note, however, that the second order coherence can be related to the order parameter only when one considers a pure coherent state. A measurement of $g_2(0)$ alone does not in general provide full information on the order parameter.

The *P*-function which describes the superposition of the thermal and coherent states reads [17]:

$$P_{\uparrow\downarrow}(\alpha) = \frac{1}{\pi n_{\uparrow\downarrow,th}} e^{-\frac{|\alpha - \alpha_{\uparrow\downarrow}|^2}{n_{\uparrow\downarrow,th}}}. \quad (10)$$

Using (10), $S(t)$ evaluates to:

$$S(t) = \frac{S(0) e^{-\left(\frac{n_{\uparrow,coh}\theta}{n_{\uparrow,th}\theta + 1} + \frac{n_{\downarrow,coh}\theta^*}{n_{\downarrow,th}\theta^* + 1}\right)}}{\left(n_{\uparrow,th}\theta + 1\right)^2 \left(n_{\downarrow,th}\theta^* + 1\right)^2} \quad (11)$$

with $\theta = 1 - \exp[-i\omega t]$ and $\omega = V/\hbar$.

We now consider the likely configuration where the *coherence degree* of spin-up and spin-down condensates are equal and given by $\chi$. In the limit $\omega t \ll 1$, expression (11) is approximately given by:

$$S(t) = S(0) \frac{e^{-in_0\rho_c\chi\omega t} e^{-\frac{1}{2}(n_0 + (1-\chi)(1+\rho_c^2)n_0^2)\chi\omega^2 t^2}}{(\frac{1}{2}(1-\chi)(1+\rho_c)n_0 i\omega t + 1)^2 (\frac{1}{2}(1-\chi)(1-\rho_c)n_0 i\omega t - 1)^2}, \quad (12)$$

The behavior of the pseudospin is dominated by the numerator of (12) in the vicinity of the coherent case ($\chi \approx 1$) and by the denominator in the opposite limit, close to the thermal case ($\chi \approx 0$). In a narrow region close to full-coherence the pseudospin oscillates in time with a period $T_0 = \dfrac{2\pi}{\omega n_0 \chi |\rho_c|}$.

Conversely to the period of oscillations, the amplitude is very sensitive to the coherence degree. The pseudospin decays like $\exp(-t^2/\tau^2)$ with characteristic time

$$\tau = \frac{\sqrt{2}}{\omega\sqrt{(n_0 + (1-\chi)(1+\rho_c^2)n_0^2)\chi}} \quad . \tag{13}$$

The decay is caused by the energy broadening of the state is induced by the huge thermal fluctuations in particle number which result in fluctuations of energy and hence on destructively interfering oscillations of the Larmor precessions. For the completely coherent case where the fluctuations in the particle number are as small as allowed without squeezing, the decay time is as high as $\tau_{coh} = \sqrt{2}/\omega\sqrt{n_0}$. If the size of the system increases at the constant density of polaritons which is so-called thermodynamical limit, $\sqrt{n_0}$ and the life-time increase, which fits well with the classical picture of Bose-condensation. However, the presence of even a tiny thermal fraction dramatically reduces the decay time. If $(1-\chi)n_0 \gg \sqrt{\chi n_0}$ the decay time estimates to $\tau = \sqrt{2}/\omega n_0 (1-\chi)$ and almost vanishes.

For values of $\chi$ below 85%, there are no oscillations and the decay is very fast and almost independent on the coherence degree. In the limit of small $\chi$, the pseudospin decays like a Lorentzian:

$$S(t) \approx \frac{1}{1 + 2i(1-\chi)n_0 \omega |\rho_c| t} \quad . \tag{14}$$

Figure 1 displays the decay time $\tau$ as a function of the coherence degree $\chi$. Parameters are those of a CdTe cavity with $n_0 = 10^4$ polaritons in the ground state and $\rho_c = 1/2$. The solid-dotted line results from numerical calculations with (12), estimating the typical lifetime as the time it takes for $|S(t)|$ to decrease by a factor $e$. The solid line superimposed is the analytical approximation (15)

4which holds over half the defining domain of $\chi$. The curve is displayed dotted below 50% where it looses physical meaning. Past this point, the decay looses its exponential character to behave according to the denominator of (13), i.e., approximately like a Lorentzian. The insets show the decay of the pseudospin in these two opposite regimes. Left inset displays the pure coherent case, where many oscillations are sustained for as long as several nanoseconds even though there is a very large number of particles. The spontaneous decay time is even longer (few hundreds of nanoseconds in the present case). It is interesting to compare the dephasing time and the typical coherence buildup time. This latter quantity is the characteristic time needed for a coherent state to appear after the non-resonant pumping is switched on. It can be calculated using kinetic theories and the typical values found for a CdTe microcavity are of the order of few hundreds of picosecond [6] which is shorter than the dephasing time of a coherent state. This comparison shows that the dephasing induced by the polariton-polariton interaction does not prevent formation of coherent states and therefore symmetry breaking in polariton systems. Right inset displays three mixed cases where the decay time has drastically decreased because of the phase mismatches brought by the thermal fraction. Since the decay time of $|S(t)|$ is very sensitive to the coherent degree of the condensate, it allows for easy and accurate measurements. Finally we point out that the dephasing time of a single component condensate can be straightforwardly extracted from the present formalism. We do not address specifically this aspect because this quantity is much harder to measure for a single component condensate than for a superposition of two different condensates.

In conclusion, we have shown that the spontaneous appearance of linear polarizaton in emission of microcavities can be considered as a criterion for Bose-condensation of exciton-polaritons, with essentials features as follows:

(i) Spontaneous symmetry breaking in a polariton system which is not fully circularly polarized should be accompanied by a buildup of linear polarization. The degree of linear polarization is dependent on the order parameters of spin-up and spin-down condensates. Its decay time increases with increase of the degree of coherence of two condensates. In case of a fully coherent state it is proportional to the square root of the number of polaritons in the condensate.

(ii) If the polariton condensate is elliptically polarized and in a coherent state, the in-plane component of its pseudospin rotates with a period proportional to the circular polarization degree of the condensate and to the number of polaritons in the ground state. This results in rotation of the main axis of the polarization ellipse of the emitted light. Thus, measuring time-resolved linearly-polarized photoluminescence one can obtain a detailed information on population of polariton condensates, their order parameters and coherence degrees.

We thank Kirill Kavokin for useful discussions. This work has been supported by the Marie Curie project "Clermont 2", contract MRTN-CT-2003-503677.

**References**

1. G. Khitrova, H. M. Gibbs, F. Jahnke, M. Kira, and S. W. Koch, Rev. Mod. Phys. **71**, 1591 (1999).

2. A. Kavokin and G. Malpuech, *Cavity Polaritons*, **ISBN:** 0-125-33032-4, Elsevier North Holland, (2003).

3. H. Deng, G. Weihs, C. Santori, J. Bloch and Y. Yamamoto, Science **298**, 199 (2002).

4. G. Malpuech, Y.G. Rubo, F.P. Laussy, P. Bigenwald and A.V. Kavokin, Semicond. Sci. Technol. **18**, S395 (2003).

5. Le Si Dang, D. Heger, R. André, F. Bœuf, R. Romestain, Stimulation of polariton photoluminescence in semiconductor microcavities, Phys. Rev. Lett. **81** 3920 (1998).

6. F. P. Laussy, G. Malpuech, A. Kavokin, and P. Bigenwald, Phys. Rev. Lett. **93**, 016402 (2004)

7. Yu. G. Rubo, F. P. Laussy, G. Malpuech, A. Kavokin, and P. Bigenwald, Phys. Rev. Lett. **91**, 156403 (2003)

8. C. Ciuti, V. Savona, C. Piermarocchi, A. Quattropani, and P. Schwendimann, Phys. Rev. B **58**, 7926 (1998).

9. D. Porras, C. Ciuti, J.J. Baumberg, and C. Tejedor, Phys. Rev. B **66**, 085304 (2002).

10. F. Tassone and Y. Yamamoto, Phys. Rev. B **59**, 10830 (1999).

11. I. A. Shelykh, A. V. Kavokin, G. Malpuech, P. Bigenwald, and F. Laussy, Phys. Rev. B **68**, 085311 (2003).

12. D. Porras and C. Tejedor, Phys. Rev. B **67**, 161310 (2003).

13. P. G. Lagoudakis, P. G. Savvidis, J. J. Baumberg, D. M. Whittaker, P. R. Eastham, M. S. Skolnick, and J. S. Roberts, Phys. Rev. B **65**, 161310 (2002).

14. M.D. Martin, G. Aichmayr, L. Viña, R. André, Phys. Rev. Lett. **89**, 077402 (2002).

15. I. Shelykh, K. V. Kavokin, A. V. Kavokin, G. Malpuech, P. Bigenwald, H. Deng, G. Weihs, and Y. Yamamoto, Phys. Rev. B **70**, 035320 (2004).

16. I.A. Shelykh, G. Malpuech, K. V. Kavokin, A. V. Kavokin, and P. Bigenwald, Phys. Rev. B **70**, 115301 (2004).

17. D. F. Walls, G. J. Milburn, Quantum Optics, Springer-Verlag (1994).

**Figure captions**

Figure 1:

Decay time of the polarization envelope (bullets on numerical points) and its analytical approximation (15) which holds in the vicinity of coherent cases (solid) for $\rho_c = 1/2$. The unphysical behavior of (15) is shown dotted. The insets show decay of the linear polarization for the full coherent case ($\chi=1$, left) featuring sustained oscillations, and the impact of thermal contamination ($\chi=0.01, 0.90, 0.99$, right). Parameters are typical for CdTe-based microcavities: $E_b$=25 meV, $a_b$=40Å, spot size 60 microns, $n_0$=$10^4$.

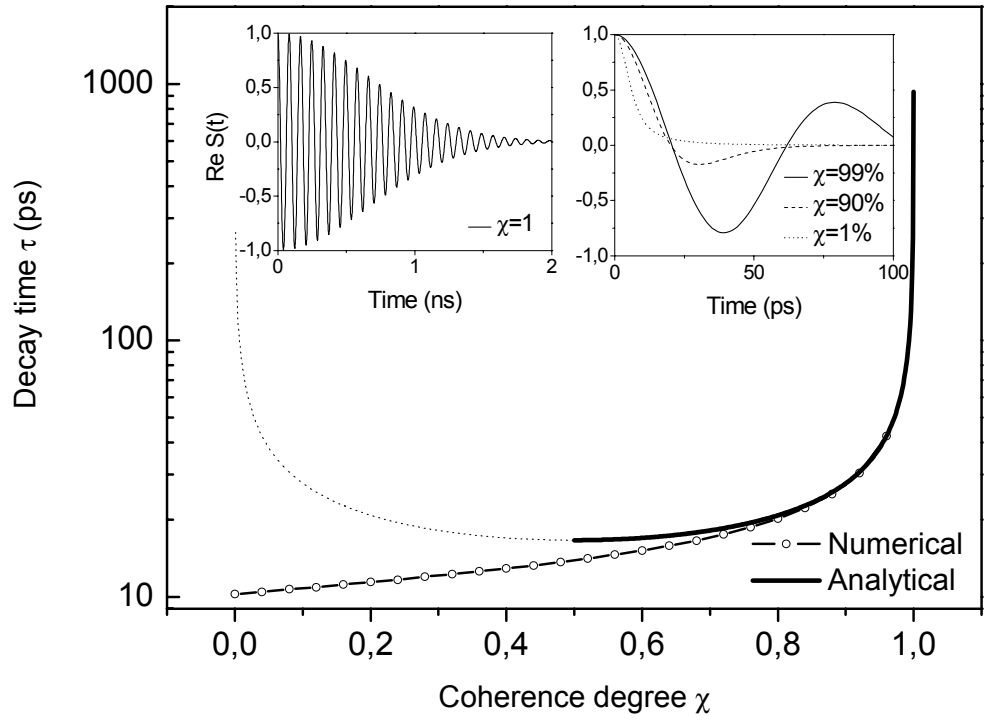